# Non-neurotoxic Nanodiamond Probes for Intraneuronal Temperature Mapping

*David A. Simpson[1,2*], Emma Morrisroe[3], Julia M. McCoey[1], Alain H. Lombard[4], Dulini C. Mendis[5], François Treussart[4], Liam T. Hall[1], Steven Petrou[2,3,6,7], Lloyd C. L. Hollenberg[1,2,8]*

[1]School of Physics, University of Melbourne, Parkville, 3010, Australia.

[2]Centre for Neural Engineering, University of Melbourne, Parkville, 3010, Australia.

[3]Florey Neuroscience Institute, University of Melbourne, Parkville, 3010, Australia.

[4]Laboratoire Aimé Cotton, CNRS, Univ. Paris-Sud, ENS Paris-Saclay, Université Paris-Saclay, 91405 Orsay, France.

[5]Department of Mechanical Engineering, University of Melbourne, Parkville, VIC 3010, Australia.

[6]Centre for Integrated Brain Function, University of Melbourne, Parkville, 3010, Australia.

[7]Department of Medicine, Royal Melbourne Hospital, University of Melbourne, Parkville, 3010, Australia.

[8]Centre for Quantum Computation and Communication Technology, University of Melbourne, Parkville, 3052, Australia.

ABSTRACT

Optical bio-markers have been used extensively for intracellular imaging with high spatial and temporal resolution. Extending the modality of these probes is a key driver in cell biology. In recent years, the NV centre in nanodiamond has emerged as a promising candidate for bio-imaging and bio-sensing with low cytotoxicity and stable photoluminescence. Here we study the electrophysiological effects of this quantum probe in primary cortical neurons. Multi-electrode array (MEA) recordings across five replicate studies showed no statistically significant difference in 25 network parameters when nanodiamonds are added at varying concentrations over various time periods 12-36 hr. The physiological validation motivates the second part of the study which demonstrates how the quantum properties of these biomarkers can be used to report intracellular information beyond their location and movement. Using the optically detected magnetic resonance from the nitrogen vacancy defects within the nanodiamonds we demonstrate enhanced signal-to-noise imaging and temperature mapping from thousands of nanodiamond probes simultaneously. This work establishes nanodiamonds as viable multi-functional intraneuronal sensors with nanoscale resolution, that may ultimately be used to detect magnetic and electrical activity at the membrane level in excitable cellular systems.

Imaging neuronal action potentials (APs) at the single cell level is a challenging and active field of research. A variety of approaches have been developed with success from patch clamping[1] and multi-electrode arrays (MEAs)[2] to optical recording *via* genetically encoded voltage indicators.[3, 4] However, many of these techniques suffer from at least one drawback of scalability, low sensitivity, slow temporal response, toxicity, low specificity and limited spatial resolution. Optical probes have attracted significant interest in recent years due to the relatively high signal to noise ratios, ~20 % increase in the fluorescence signal at the peak of the AP,[5] and the possibility of high resolution sub-micron imaging of APs in real time. Unfortunately, these approaches are often accompanied by toxicity effects due to the overexpression of the genetically encoded probe, and can in some cases modify the network activity of the system under investigation. Therefore, the development of alternative sensing probes that exhibit low neurotoxicity and can report cellular physico-chemical events such as changes in the electric potential, pH, ionic currents, magnetic field and potentially temperature is essential.

Diamond nanocrystals containing colour centres have been identified as promising candidates for bio-imaging with long term photo-stability and inherently low cytotoxicity.[6] These properties have led to a myriad of fluorescence based applications including neuronal differentiation and tracking. The presence of nanodiamonds (NDs) in such systems has been shown not to alter the cell viability,[7] morphology[8] or production of neuron-specific markers such as β-III-tubulin during cell differentiation.[9] The photo-stability of the probes has also been used to identify cellular internalization pathways,[10] and monitor and track the intraneuronal endosomal transport in model and transgenic systems.[11] The prominent optical defect centre used in these biological applications is the negatively charged nitrogen-vacancy (NV) centre. In addition to the photo-stability of this defect, the quantum properties of the NV provide opportunities to sense a range of physio-chemical parameters including magnetic fields,[12,13] temperature,[14] electric potentials[15] and pH[16] which are of particular interest in excitable cellular systems. In order to utilise these nanoscale quantum sensors in functioning neuronal networks, we must first establish whether the presence of the probe itself has an adverse effect on the connectivity and performance of the neuronal network. Neurotoxicity can often proceed in the absence of other biochemical or morphological changes;

therefore, traditional cell toxicity assays are not optimized for detecting this type of toxicity.[17] MEAs have demonstrated the capacity to screen for neurotoxic effects such as synaptic function, action potential generation and propagation, plasticity, and network formation and function.[17, 18]

Here we use MEAs to assess whether the presence of NDs in primary cortical neurons elicits a neurotoxic response. We conduct a detailed network analysis across multiple control and ND groups at varying ND concentrations. Comparison of 25 separate network performance indicators over time demonstrate no neurotoxicity from the uptake of NDs at concentrations up to 20 µg/ml. In addition to the neurotoxicity study, optical imaging and quantum sensing protocols were performed on the primary cultures. We demonstrate improved signal to background fluorescence imaging by taking advantage of the optically detected magnetic resonance (ODMR) associated with the NV centre and use wide-field optical techniques to simultaneously report the ODMR spectra from thousands of NDs within seconds. As a proof-of-principle demonstration we apply quantum control techniques to map the intracellular temperature of a neuronal network.

Action potentials which govern interactions and connectivity in excitable cells and neuronal networks generate a range of measurable parameters. To date, optical sensors have been developed to respond to changes in electrical potential and/or ionic currents. Nanodiamonds represent a promising approach for the detection of APs with demonstrated sensitivity to magnetic and electric fields, pH and temperature. At present, ND magnetic[19] and electric field[15] achieved sensitivities are far from that required to detect single AP events.[20] However, recent reports by Barry *et al.* [21] have shown that magnetic detection of single APs in squid and worm axons is possible by integrating the signal from ensembles of NV centres in single crystal diamond over large volumes ($0.013 \times 0.2 \times 2$ mm$^3$). Beyond magnetic sensing, opportunities exist based on photoluminescence (PL) changes in response to electrostatic potential modulation,[15] but it remains to be seen if the electric signaling from ion exchange across axonal membranes, as opposed to charge injection, can modulate NV fluorescence. These potential applications motivate further investigation of NDs in biological systems, in particular, excitable cellular systems where these types of signals are generated. However, before deploying such sensors into such complicated systems, it is imperative to evaluate the neurotoxicity of these probes. The motivation of this work is to assess whether the interaction of NDs with neurons impacts the functional dynamics and performance of neuronal networks and to show that the imaging protocols required for quantum sensing can be achieved on a timescale relevant to cellular processes.

RESULTS

**Multi-electrode array recordings of neuron primary cultures containing fluorescent NDs.**

Here we use MEAs to systematically study the effect of NDs on the electrophysiological activity of cultured mouse cortical neurons between DIV 11 and DIV 13. Extracellular electrical recordings were carried out using the Multiwell Screen from Multi Channel Systems in 5 minute recordings at 20 kHz. During data acquisition, cultures were kept at 37 °C and enclosed in chambers with perfused carbogen (95% $O_2$, 5% $CO_2$). The network electrical activity was benchmarked with control cultures injected with commensurate amounts of sterile water. Figures 1(a-d) show typical AP spikes from a single MEA channel over a 10 s period, for the control and ND-containing networks. The network parameters across the 12 channels per well can be visually represented in a raster plot as shown in Fig. 1(e-h), where each vertical line represents a single AP and each horizontal row represents one channel of the 12 in each MEA well. From this plot, it is possible to visualise spikes, bursts, and network bursts across each channel. The raster plots for the baseline control and ND groups show a high level of network bursting, consistent with a relatively mature neuronal network. Five sets of MEA recordings were performed on 48 individual wells at varying ND concentrations. The first set of MEA baseline recordings were undertaken at Day 11, prior to the introduction of the NDs. The control recordings were performed 24 hours post media change to avoid stress-related responses. The ND suspension was dispersed in cell media, sonicated for 5 min and then applied to the primary cultures during a routine change of cell media. Network recordings were then carried out at 12, 24 and 36 hours intervals.

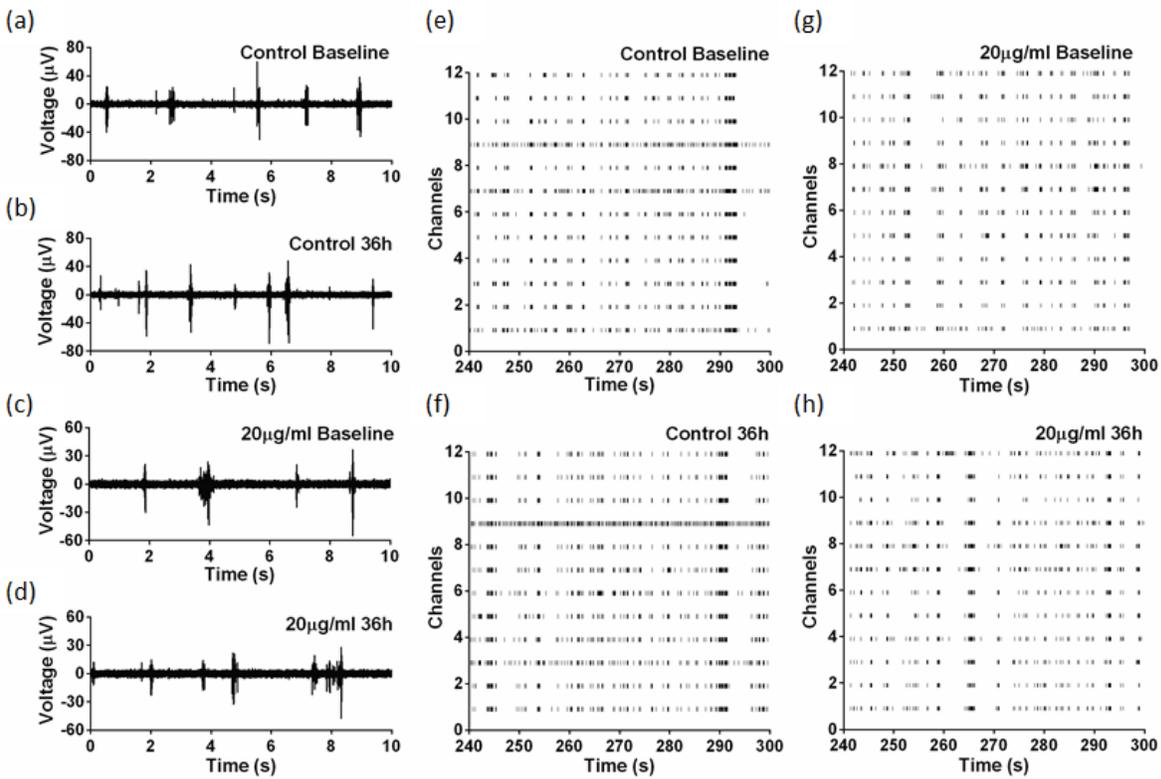

**Figure 1**: Electrical recordings obtained from the Multiwell Screen for two conditions at baseline (a) and (c) and 36 hours after the addition of water vehicle (b) and 20 µg/ml ND (d). The data signals have been high-pass filtered at 300 Hz to detect AP spikes. (e-h). Raster plots showing AP spikes from each of the 12 electrodes per well for the baseline, control and highest concentration ND group. The plots are shown for the last minute of recording.

To quantify the level of network activity, key parameters of the network spiking such as the burst rate, firing rate, spike amplitude and percentage of spikes in network bursts are compared from all channels across the 48 individual wells for the control and ND groups. The results for the baseline and highest ND concentration (Fig. 1) are typical for the results observed across all measured ND concentrations. The raw comparisons of four specific network parameters are shown in Figure 2(a-d). The plots show no statistically significant difference in any of the key network comparisons both in terms of absolute values and trend across the data set. In all cases, the rate of network bursting increased with time as did the mean firing rate and average spike amplitude. This provides strong evidence that the network is continuing to mature and strengthen its connectivity across computational units, even after the addition of NDs. The highly synchronous nature of the activity and the fact that greater than 80 % of spikes occur in a network burst indicates highly interconnected neurons within the network. A more detailed network comparison is shown in Figs. 2(e) and (f). For each parameter, the percentage change from the baseline was calculated at each time point. Any differences between a ND group and the corresponding control that are statistically significant would be coloured while those that are not significant will be shown in grey.

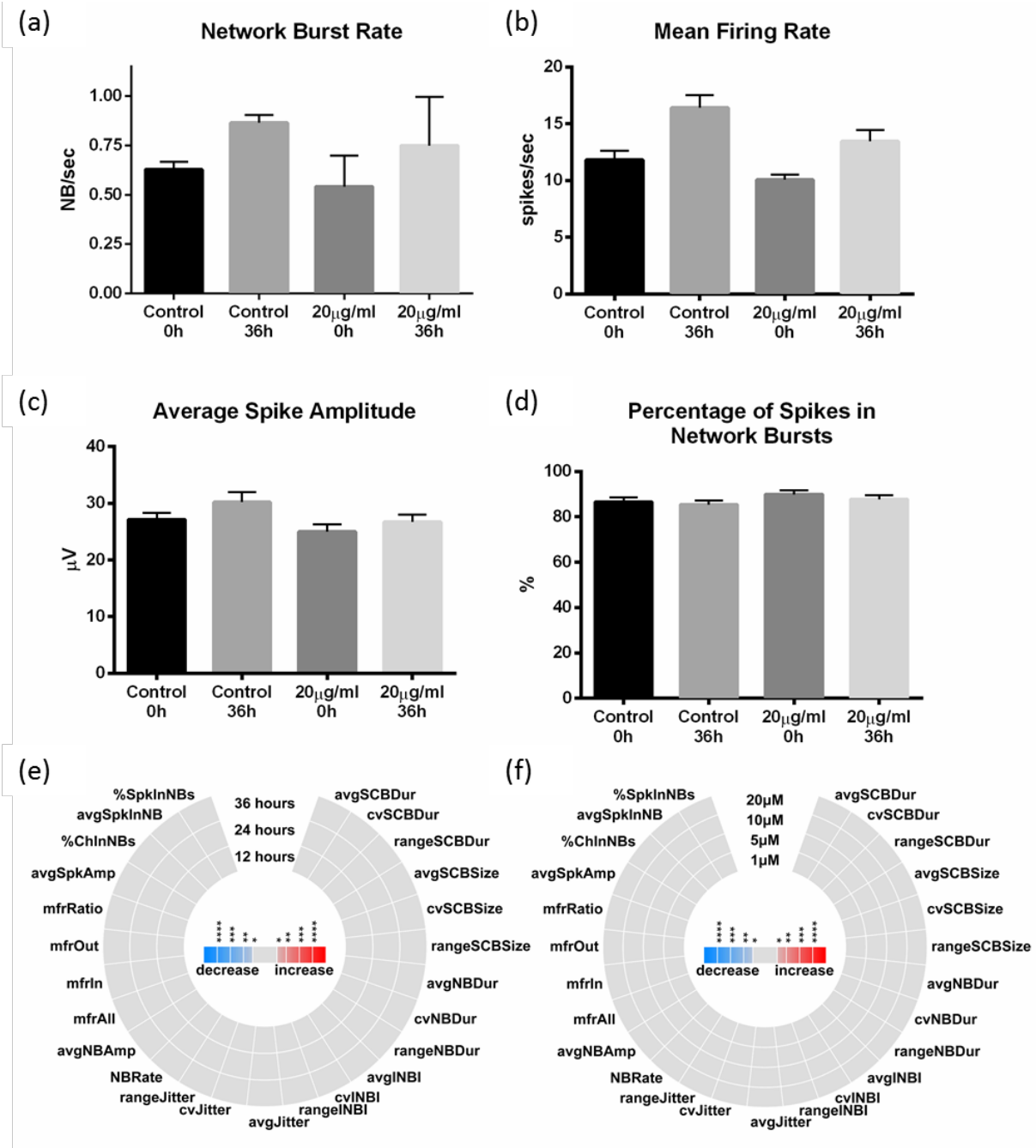

**Figure 2**: Neuronal network parameter comparison for varying ND concentrations. (a-d) The network burst rate, mean firing rate, average spike amplitude and percentage of spikes in the network bursts for the control and highest ND concentration (20 µg/ml) at 0 hours and 36 hours across the 5 replica sets. (e-f) Iris plots comparing 25 network parameters between control and ND groups. Grids in each heat map represent statistical comparison between a ND condition (e) 12, 24, 36 hours after the application of 20 µg/ml NDs,

(f) 36 hr after the application of 1, 5, 10, 20 µg/ml concentrations of NDs and its control. Sections in the circles correspond to network parameters. *p*-values resulting from statistical comparisons (adjusted for multiple comparisons) are represented by the colour scale, with grey showing no difference between the pair of conditions being compared. Red and blue represent *p*-values for increase/decrease in the parameter values in the ND group compared to control as indicated within the iris plots (* $0.01 \leq p < 0.05$, ** $0.001 \leq p < 0.01$, *** $0.0001 \leq p < 0.001$, **** $0.00001 \leq p < 0.0001$).

As can be seen from the iris plot, there is no statistically significant difference in 25 separate network comparison parameters investigated over time or as a function of ND concentration. This detailed comparison demonstrates that NDs present at concentration ranges up to 20 µg/mL (1.6 µM) have a negligible impact on the neuronal activity, growth or development. The next section demonstrates how the quantum properties of NV-containing NDs can be optically addressed, manipulated and read out simultaneously over wide fields of view. As a proof of principle application, we show how to report the local intracellular temperature using the quantum properties of the probes.

**Wide-field imaging and ODMR spectroscopy of NDs.**

The optical and quantum properties of the NV centre in diamond set it apart from other fluorescence based probes. These properties can be controlled and manipulated at room temperature in biological environments[22] to report changes in the local pH, electric, magnetic and temperature fields. Over the past decade there has been a renewed interest in probing the intracellular temperature with several fluorescence based methods proposed to image the temperature distribution including: fluorescence proteins[23] and polymers,[24] quantum dots,[25] rare-earth doped nanocrystals,[26] metal nanoparticles[27] and NDs.[14] Many of these approaches rely on fluorescence changes in the excited state lifetime or the fluorescence spectrum of the probe. The temperature response from NDs differs from most fluorescence probes, whereby the ground state spin levels are shifted in response to a change in temperature.[28] This provides a robust readout of temperature with reduced cross sensitivity to other physiological parameters such as pH and redox potential. Therefore, in this section we demonstrate how these nanoscale quantum probes can be used to report intracellular temperature distributions over wide-fields of view within a temporal resolution of order a few seconds. To demonstrate optical and quantum sensing protocols in primary cortical neurons we culture neuronal networks onto a tailored glass coverslip with a gold microwave resonator coated with polydimethylsiloxane (PDMS) using the same culturing procedure outlined for the MEA investigation (see Methods). The ND suspension was dispersed in cell media (6 µg/ml), sonicated for 5 min and then applied to the primary cultures during a routine change of cell media. Imaging was carried out 12 hours after the incubation. Using ODMR spectroscopy[29] in combination with standard wide-field microscopy techniques, see Fig. 3(a), we are able to simultaneously determine the ground state energy level structure from thousands of individual NDs over a wide field of view (80 × 80 µm) within seconds. The ground state spin levels of the NV centre respond to local magnetic, electric and temperature fields in a variety of ways. For example, in the presence of a magnetic field, the spin states Zeeman split with a gyromagnetic ratio of 2.8 MHz/G.[12] An electric field Stark splits the energy levels by 1.7 kHz m/V,[30] whilst changes in temperature shift the spin levels by -74 kHz/K.[28] In addition to these effects, the ODMR provides a mechanism to modulate the photoluminescence from NV centres within the NDs.[31] At zero magnetic field, the majority of NDs have a crystal field splitting of $D = 2.87$ GHz. Under 532 nm laser excitation, the NV spin can be conveniently polarised into the $m_S = 0$ bright state. By applying a microwave frequency of 2.87 GHz across the field of view, the NV spins can be driven into the $m_S = \pm 1$ dark state with a photoluminescence (PL) ratio $\Delta PL/PL$ of ~10 %. By taking the difference of two photoluminescence images with and without the microwave field as shown in Fig. 3(b) unwanted cell/media auto-fluorescence can be removed[31]. This is particularly important for neuronal cell imaging which often requires tailored cell culture media along with well controlled temperature environments to maintain functional networks. PL images of primary cortical neurons incubated with NDs at a concentration of 6 µg/mL of cell media are shown in Fig. 3(c), along with alternative ODMR contrast images of the same area obtained with microwaves "on" and "off" resonance at 2.87 GHz and 2.84 GHz, respectively. By comparing a single diffraction limited spot from an individual ND in the PL and ODMR contrast image, we find an 50% increase in the signal to background ratios for the ODMR contrast image due to the removal of cell and media auto-fluorescence. The ODMR contrast imaging for the "off" resonance case shows no fluorescence contrast from the same imaging area demonstrating the ODMR contrast enhancement acts over the entire field of view and is highly selective.

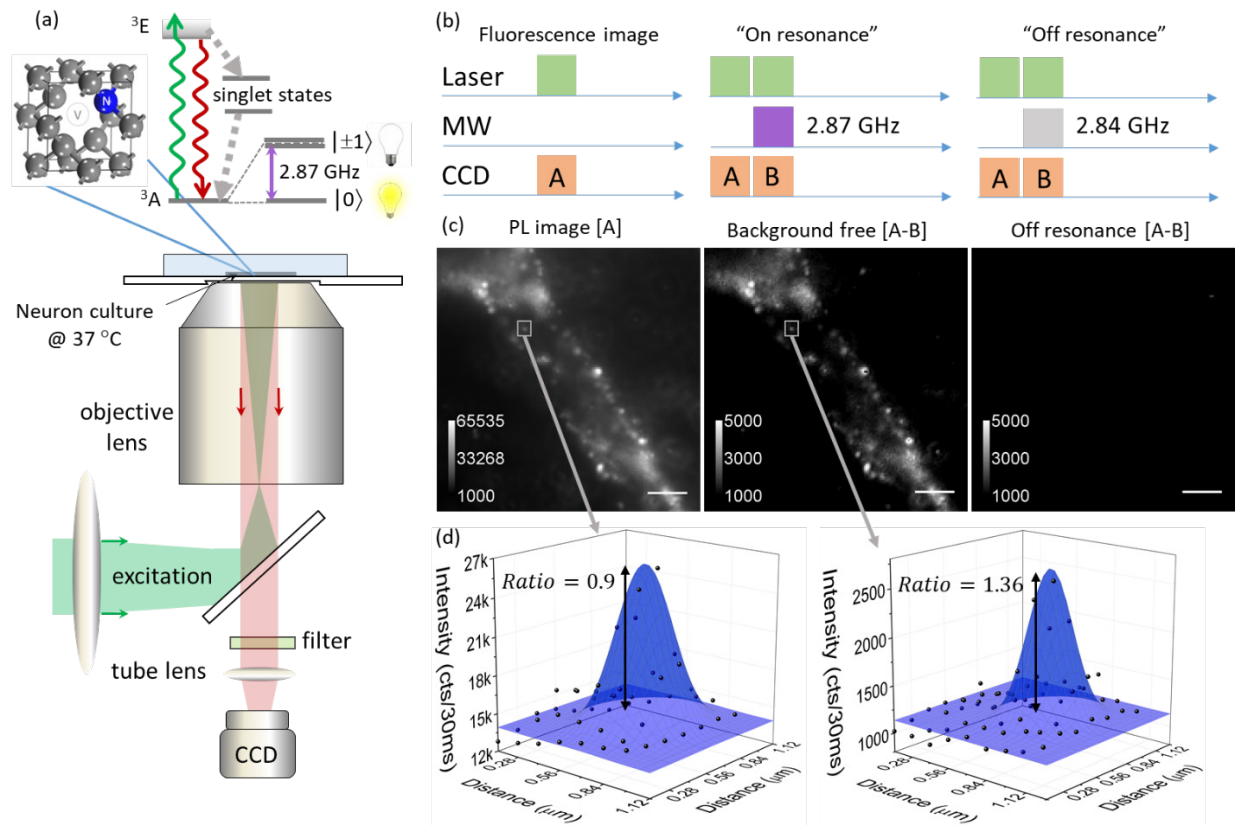

**Figure 3**: Wide-field imaging of NDs containing NV centres in primary cortical neurons. (a) Structure and energy levels of the NV centre in diamond and schematic of the wide-field fluorescence microscope. A 532 nm laser is used to excite the NDs with the resulting fluorescence collected from 650-750 nm. (b) Acquisition sequences for conventional PL and ODMR contrast imaging. ODMR contrast imaging is performed "on" (2.87 GHz) and "off" (2.84 GHz) resonance with a microwave pulse applied over the field of view. (c) Series of images using the acquisition sequences described in (b). The PL image (left) shows NDs distributed throughout the cellular components. The ODMR contrast image (middle) "on" resonance shows background free imaging with improved signal to noise ratios, while the "off" resonance image (right) demonstrates the microwave pulse acts efficiently over the entire field of view. (d) 3D fluorescence intensity maps for a single diffraction limited ND spot highlighted in (c). Analysis shows an 50% increase in the signal to background ratio for the ODMR contrast imaging technique.

The entire ODMR spectrum from each individual ND can be acquired through an image series at various microwave frequencies. The wide-field NV fluorescence image in Fig. 4(a) taken 6 μm above the coverslip identifies thousands of NDs within the neuronal structures. Figure 4(b) shows typical ODMR spectra from two individual NDs from Fig. 4(a), with a typical ODMR peak splitting observed due to intrinsic strain within the ND. The ODMR spectrum is obtained from a set of 100 images integrated for 30 ms per image with a total acquisition time of 6 seconds. Although each individual ND may experience slightly modified crystal field splitting $D$ from local strain,[22] the 170 nm NDs used in this work exhibit a spread in $D$ smaller than 170 kHz measured over 9078 different NDs which ensures the ODMR contrast imaging technique is effective in addressing all fluorescent NDs at zero magnetic field.

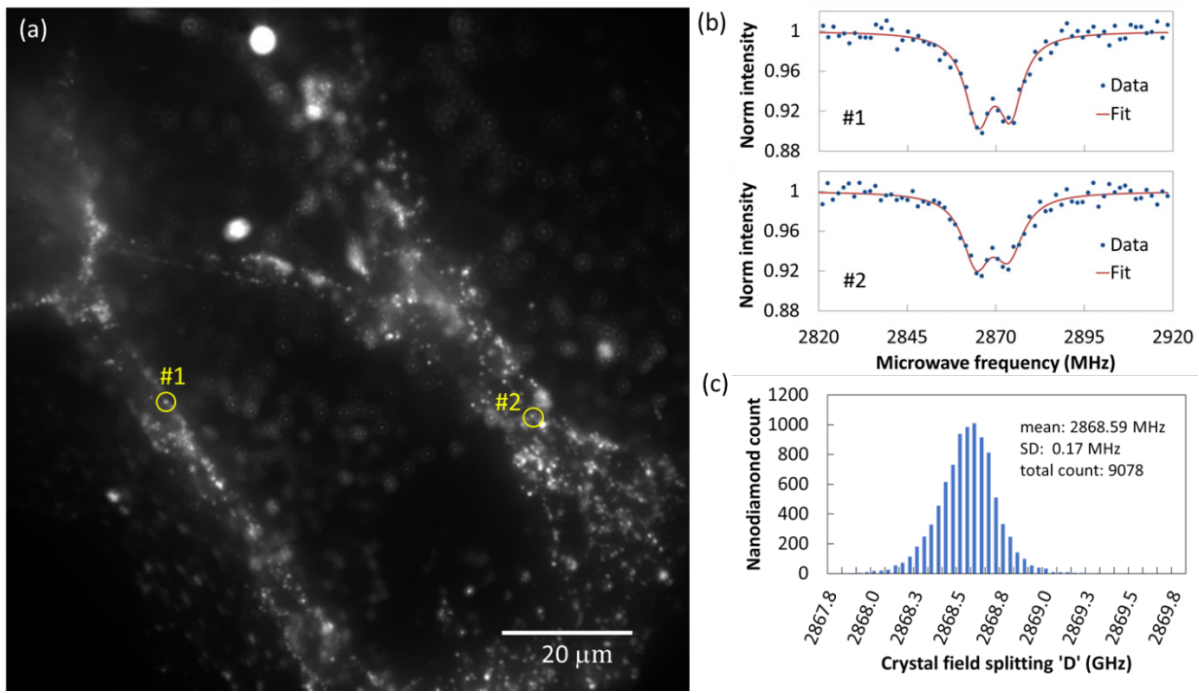

**Figure 4**: Wide-field ODMR imaging of NDs. (a) NV fluorescence image of NDs measured 6 μm above the coverslip, recorded at 37 °C. (b) ODMR spectra from two individual NDs corresponding to the points #1 and #2 in (a). The ODMR spectra is fit with a sum of two Lorentzians (red line). The total acquisition

time for the simultaneous recording of ODMR spectra of all NDs was 6 seconds. (c) Histogram of the distribution of *D* from 9078 NDs in Fig. 4(a). The mean crystal field splitting was 2868.59 MHz with a standard deviation of 170 kHz.

The ND concentration used for imaging (6 µg/mL) provides good coverage of the network and may be used to spatially map intracellular components. To demonstrate the sensing capabilities of the ND probes, the temperature of the environmental chamber was reduced by 1.9 °C. A temperature map from the intracellular NDs, shown in Fig. 5a, was obtained by identifying individual ND spots in the fluorescence images at 37.3 and 35.2 °C (see Methods for procedure). The spots were then co-localized within a maximum pixel distance of *N*=9, corresponding to 2 µm. This analysis resulted in 255 co-localised regions from which the ODMR spectrum was obtained at each temperature. The crystal field splitting, *D*, from each ND was determined by fitting a single Lorentzian peak to the ODMR spectrum, in place of the two Lorentzian peak fit shown in Fig. 4b. The noise around the strain split ODMR lines causes significant uncertainly in the crystal field splitting parameter. We instead remove 4 central points and fit a single Lorentzian to the data to achieve a more precise estimate of *D*, as shown in Fig. 5b. The *D* values from the co-localised NDs were retained for ODMR fits with errors in the peak position less than 0.01%. A temperature map was then obtained by subtracting the *D* at each location at 37.3 and 35.2 °C and converting the shift in *D* into temperature, using $dD/dt$ = -74 kHz/K. The acquisition time per point was 120 ms with 50 points in total used for the fitting analysis. With pixel binning of 1.2×1.2 µm and a total acquisition time of 12 s the signal to noise ratio for the temperature recording just above one. Figure 5(b) shows an example of the ODMR fit from a single co-localised spot with a shift in *D* of 110 ± 90 kHz and a change in local temperature of -1.5 ± 1.2 °C. The uncertainty in *D* is dominated by the broad ODMR line associated with the noise from the nitrogen and $^{13}$C spin bath intrinsic to the ND. The precision can be improved using the statistical power of many NDs as shown below or by utilizing tailored ultra-pure NDs currently under development.[32] Temporal improvements can be made by moving to a simplified two[33] or four point[34] frequency measurement technique and increasing the pixel binning at the expense of spatial resolution. A histogram of the temperature recordings from Fig. 5(a) is shown in Fig. 5(c). The distribution reports a mean temperature change of -1.36 °C with a standard error on the mean (s.e.m.) of 0.08 °C. This is consistent with the reduction in environmental temperature measured externally from the network culture. The measurement technique was verified by performing the same fitting procedure on two sets of fluorescence images at the same environmental temperature *i.e.* 37.0 ± 0.1 °C. The histogram from 209 reported NDs found a mean temperature change of 0.2 ± 0.1 (s.e.m.) °C, see Supplementary Information.

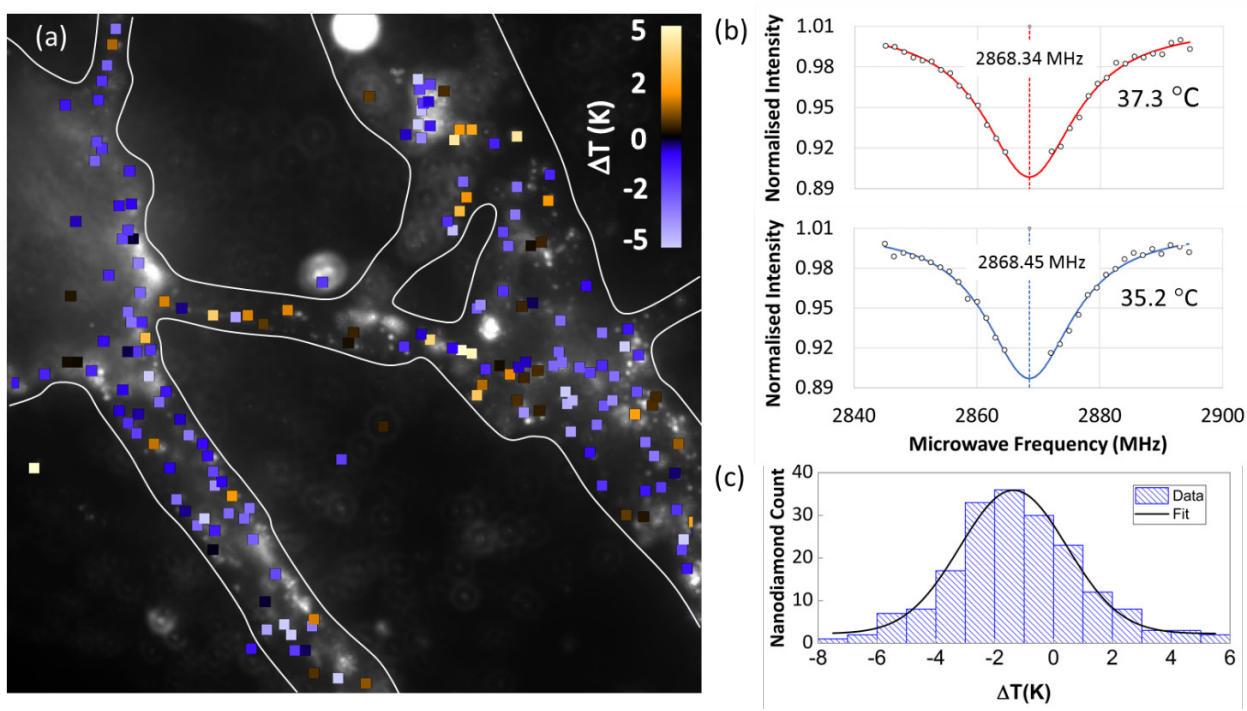

**Figure 5**: Intracellular temperature map from NDs in primary cortical neurons. (a) Temperature map from co-localised NDs overlaid with the ND fluorescence image at 37.3 °C. (b) ODMR spectra from a typical

ND in (a) at 37.3 °C (top right) and 35.2 °C (bottom right). The shift in $D$ of $110 \pm 90$ kHz represents a local temperature change of $-1.5 \pm 1.2$°C. The temperature error is obtained from the square root of the sum of the square errors on the peak position from the respective fits at each temperature. (c) Histogram of temperature changes measured from 255 co-localised NDs. The mean temperature change reported across the neuronal network was $-1.36 \pm 0.08$ (s.e.m) °C consistent with the reduction in environmental temperature.

This proof of principle demonstration showed no functional or morphological neuronal degradation from the weak optical and microwave fields used in the temperature sensing protocol, indicating that NDs can act as effective multi-modal sensors in functioning neuronal networks. Recent studies have identified temperature heterogeneity in neuronal cells,[24] however debate remains about whether biochemical reactions at the single cell level are sufficient to generate such heterogeneity.[35, 36] Other forms of temperature variations have been proposed which may accompany current flow in open ionic channels [37]. NDs therefore represent a realistic pathway towards characterising such changes at these boundaries.[14] The sensitivity of the ND probes *in vitro* is $1K/\sqrt{Hz}$; this is favorable in comparison to other nanoscale probe techniques.[14] The accuracy of the temperature recordings can be improved to less than 1 K with modest integration times of tens of seconds. In terms of the potential for magnetic and electric field sensing with these probes, the sensitivity remains limited by the quality of the diamond material. At present, the bulk of ND material arises from low quality high pressure high temperature synthesised diamond. The relatively high nitrogen (100 ppm) and $^{13}C$ (1 ppt) content in this material severely impacts the linewidths of the ODMR spectrum to of order MHz. This is three orders of magnitude away from that needed to detect single action potential events. With material improvements, such as tailored CVD growth[32] and isotopically enriched $^{12}C$ diamond[38], combined with targeted functionalisation strategies, these functional quantum probes may provide sufficient sensitivity to evaluate intracellular processes of excitable cells, with nanoscale resolution[20].

DISCUSSION

In this work, we have shown that NDs containing NV centres present a promising pathway for multi-modal imaging in biology. We have shown that the presence of NDs in neuronal networks up to a concentration of 20 µg/mL has a negligible impact on the network dynamics and connectivity. Network electrophysiological parameters such as the mean firing rate, network burst rate and spike amplitude show no statistically significant deviation from the measured controls for the five replicates at each ND concentration studied in this work. Furthermore, we demonstrate that ODMR spectroscopy can be performed in the intracellular environment using wide-field microscopy, with quantitative ODMR spectra obtained in seconds. The measurement protocol is the basis of the majority of quantum sensing protocols of the NV centre in diamond. We use the ODMR signature from individual NDs to demonstrate intracellular temperature mapping with high spatial resolution. This approach may be of interest for measuring temperature changes in response to external stimulus such as light. Optogenetics is a well-established and widely used technique for neuronal stimulation. The heat generated from the optical stimulation has been found to impact the firing dynamics of the system under investigation[39], however, quantifying these effects is challenging and has so far relied on the use of thermistors[39]. The ND-based thermometry demonstrated here appears ideally suited to this task. The temperature sensitivity of the NDs is sufficient for this type of application and the optimal NV excitation wavelength (560 nm) is distinct from the one of the most common light-gated ion channel proteins (e.g. channel rhodopsin ChR2, absorption maximum at 460 nm wavelength). Moreover, we have shown that NDs do not elicit a neurotoxic response *in vitro*. Other future applications may include the study of signaling cascade resulting from locally induced temperature changes. For example, in neurological disorders such as epilepsy, where the firing rate of neurons is found to increase; this may result from a local increase in temperature related to increased metabolism. Our method provides the opportunity to correlate temperature and neuronal activity changes. ODMR is certainly not the only spectroscopy and sensing methodology applicable to NDs; the coherence of the NV spin can also be exploited for detection of electronic[40-43] and nuclear spin[44-47] species. Therefore, applying these functional quantum probes to the intracellular environment of excitable cells may provide insight into the mechanisms and processes governing intracellular dynamics and communication.

METHODS

**Culture Preparation.**

Multiwell MEAs with gold electrodes and an epoxy (FR-4) substrate from Multi Channel Systems (Reutlingen, Germany) with a 4×6 wells format were used. MEAs were plasma cleaned 4-7 days prior to plating and were treated with PolyEthyleneImine 24 hours before plating. Wells were coated with Laminin 2-3 hours before plating and kept in the incubator. Laminin was removed immediately prior to plating. The cortex was dissected from C57BL/6 mice 0-2 days post-natal. Cell media was prepared with 89.3 % Minimum Essential Medium (MEM), 0.9 % 1M Hepes, glucose (6 mg per 1 mL of MEM), 8.9 % Fetal Bovine Serum (FBS) and 0.9 % P/S. Cortical pieces were subjected to dissociation using Trypsin (0.25 %), DNase (0.032 %) and trituration with a glass pipette. 375,000 cells were plated in 120 µL of media to cover the base of each well and after the initial media was removed, a 500 µL of culture media containing Neurobasal-A medium supplemented with 1.9 % B27, 0.95 % GlutaMax, 0.95 % HEPES and 0.95 % PenStrep was added. MEAs were covered with lids and kept in an incubator (37 °C, 65 % humidity, 9 % $O_2$, 5 % $CO_2$). 5 µM Ara-C was added at DIV 3 and was removed on DIV 5 with a 100 % medium change to control the proliferation of glial cells. 100 % culture medium changes were carried out three times a week. All animal experiments were approved by the Howard Florey Institute Animal Ethics Committee, and performed in accordance with the Prevention of Cruelty to Animals Act and the NHMRC Australian Code of Practice for the Care and Use of Animals for Scientific Purposes.

**MEA Analysis.**

Voltage data recorded with Multiwell Screen system (Multi Channel Systems) were imported into Matlab (The MathWorks, Inc., Natick, Massachusetts, USA) where subsequent analysis was carried out. In order to extract high frequency components of the signals, voltage signals of each electrode were high-pass filtered at 300 Hz. Spikes were detected with custom Matlab scripts based on precision timing spike detection,[48] with spikes being detected if the highest phase of the spike was greater than 6 times the standard deviation of noise. Bursts on single channels were detected depending on average firing rates as described in Mendis *et al.*[49] with the minimum number of spikes for detecting a network burst set at three. Network bursts (NBs) were detected when more than 25 % of channel-wise bursts overlapped in time.

Network characteristics were extracted in terms of firing and bursting parameters. Average amplitude of spikes (avgSpkAmp) and mean firing rates (MFRAll) were calculated for each channel over the entire time period and were then averaged across channels. Mean firing rates inside bursting periods (MFRIn), in non-bursting periods (MFROut) as well as the ratio MFROut/MFRIn (MFRRatio) were calculated. Burst features were calculated for both single-channel bursts as well as network bursts. Single-channel burst parameters include the number of spikes inside bursts (SCBSize) and burst durations (SCBDur). Network burst parameters features consists of NB rate, NB durations (NBDur), inter NB intervals (INBI-time interval between the end of a NB and the beginning of the next NB), the jitter between channels participating in the NB (time interval between the starts of the first and last single-channel bursts inside a NB), the average number of spikes in a NB (avgSpkInNB), the amplitude of spikes in a NB averaged over the NB duration (avgNBAmp), the percentage of active channels participating in NBs (%ChInNBs) and the percentage of spikes in NBs (%SpkInNBs). For the parameters SCBSize, SCBDur, NBDur, INBI and jitter, the mean, coefficient of variation (cv) and the range were calculated.

**Statistical analysis.**

The statistical differences for each network parameter was characterised by looking at the percentage change from the measurement baseline between the control and ND groups. If the values in each group followed a Gaussian distribution (as determined by Jarque-Bera tests), *t*-tests were carried out to obtain the statistical difference between them. Otherwise, Wilcoxon Mann Whitney-U tests were carried out. The resulting *p*-values were corrected for multiple comparisons using the Bonferroni method and were used to generate iris plots, which are radial heat maps that show color-coded *p*-values. Lower *p*-values (ex: $p < 0.0001$) represent a larger difference between the two groups while higher *p*-values (ex: $0.01 < p < 0.05$) represent smaller differences.

**ODMR fitting procedure.**

The majority of NDs exhibit strain splitting in their ODMR spectrum as shown in Fig. 4(b). To improve the peak fitting precision, the four central data points were removed and a single Lorentzian fit applied. For the temperature mapping, NDs were located using a spot detection algorithm using the open source software package Icy. Bright spots were detected over a dark background with a lower scale of: 100 and an upper scale of: 700. Nanodiamond locations were then co-located using the "Co-localisation studio" package for the respective temperatures of 37.3 and 35.2 °C with a maximum pixel distance of $N=9$. The ODMR peak position was then obtained from a 6×6 bin area around each co-localised spot. The data was retained for Lorentzian fits with less than 0.01% uncertainty in the peak position. The change in peak position at each location was then converted into temperature using the relationship $dD/dT = -74$ kHz/K.

**Materials.**

The NDs (brFND-100) used in this work were sourced from FND Biotech Inc. (Taiwan). The NDs were dispersed in water at a concentration of 1 mg/ml. The Z-average particle size from dynamic light scattering measurements was 170 nm (see supplementary information) and the average number of NVs per particle was ~500. No specific ligands/moieties were used to facilitate cellular uptake. The uptake was passive and nonspecific endocytosis.

**Fluorescence Imaging.**

The widefield imaging and quantum sensing experiments were undertaken by preparing neuronal cultures onto a glass coverslip with an evaporated gold microwave resonator. The glass coverslip was coated with a thin layer of PDMS (10:1 ratio) to prevent contact between the active and ground plane of the resonator when the cultures were immersed in cell media. The networks were cultured under the same conditions as described above. The ND suspension (brFND) was dispersed in cell media (6 µg/ml), sonicated for 5 min and then applied to the primary cultures during a routine change of cell media. Imaging was carried out 12 hours after the incubation. The wide-field imaging was performed on a modified inverted microscope (Eclipse Ti-U, Nikon, Japan). Optical excitation from a diode-pumped solid-state laser emitting at 532 nm wavelength (Sprout G, Lighthouse Photonics, Inc, San Jose, USA) was focused onto an acousto-optic modulator (AOM) (Model 3520-220 from Crystal Technologies/ Gooch & Housego, Palo Alto, USA) and then expanded and collimated (GBE05-A beam expander, from Thorlabs Inc, Newton, USA) to a beam diameter of 10 mm. The collimated beam was focused using a wide-field lens (focal length: 300 mm) to the back aperture of the Nikon x40 (1.2 NA) oil immersion objective *via* a dichroic mirror (Ref. Di02-R561-25×36, Semrock Inc, Rochester, USA). The NV fluorescence was filtered using two bandpass filters before being imaged using a tube lens (focal length: 300 mm) onto a sCMOS camera (Neo, Andor Technology Ltd, Belfast, UK). The optical configuration relies on a standard widefield optical microscope using laser light excitation source, with the addition of the AOM for optical modulation. Microwave excitation was provided by a microwave generator (Ref. N5182A, Agilent Technologies, Santa Clara, USA) and switched using a RF switch (Ref. ZASWA-2-50DR+, Miniciruits Inc, Brooklyn, USA). The microwaves were amplified (Ref. 20S1G4, Amplifier Research Corp., Bothell, USA) before being sent to the microwave resonator. A computer board (Ref. PulseBlasterESR-PRO 500 MHz, SpinCore Technologies, Inc, Gainesville, USA) was used to control the timing sequences of the excitation laser, microwaves and sCMOS camera and the images where obtained and analysed using custom LabVIEW code. Optical imaging was performed at 37 °C unless stated otherwise with an excitation power density of 30 W/mm$^2$. The temperature studies were undertaken by reducing the temperature of the environment and monitoring these changes with a k-type thermocouple and reader (Fluke Corp., Everett, USA).

ASSOCIATED CONTENT

The Supporting Information is available free of charge on the ACS Publications website at DOI: XXXXX. More information on the temperature measurement protocol are provided in the Supplementary Information, along with dynamic light scatting and zeta potential characterization of the NDs used in this work (PDF).

AUTHOR INFORMATION

**Corresponding Author**

*simd@unimelb.edu.au.

**Author Contributions**

D.A.S., E.M., S.P., L.C.L.H conceived the study. D.A.S. and E.M. performed the electrical recordings. E.M cultured the cortical neurons for the multi electrode array recordings and imaging experiments. D.A.S., J.M.M. and A.L. performed the optical imaging and quantum sensing experiments. D.A.S. constructed the quantum imaging microscope and acquisition software. D.C.M. performed the statistical analysis of the electrical recordings. D.A.S., L.T.H. and F.T. analyzed the data from the quantum measurements. D.A.S., S.P. and L.C.L.H. supervised the project and all authors contributed to writing the manuscript.

ACKNOWLEDGMENT

The authors acknowledge Prof Michel Simonneau for helpful discussions regarding applications of the technology. Dr Philipp Reineck for performing the DLS and Zeta potential measurements on the ND solutions and Dr Babak Nasr for the helium ion microscope images. This research was supported in part by the Australian Research Council Centre of Excellence for Quantum Computation and Communication Technology (Project number CE110001027). This work was also supported by the University of Melbourne through the Centre for Neural Engineering and the Centre for Neuroscience. L.C.L.H acknowledges the support of the Australian Research Council Laureate Fellowship Scheme (FL130100119). S.P acknowledges the support from the NHMRC fellowship scheme (1005050). D.A.S acknowledges support from the Melbourne Neuroscience Institute Fellowship Scheme.REFERENCES

1. Mathias, R. T.; Cohen, I. S.; Oliva, C., Limitations of the Whole Cell Patch Clamp Technique in the Control of Intracellular Concentrations. *Biophys. J.* **1990,** *58*, 759-770.
2. Spira, M. E.; Hai, A., Multi-Electrode Array Technologies for Neuroscience and Cardiology. *Nat. Nanotechnol.* **2013,** *8*, 83-94.
3. Knöpfel, T., Genetically Encoded Optical Indicators for the Analysis of Neuronal Circuits. *Nat. Rev. Neurosci.* **2012,** *13*, 687-700.
4. Gong, Y.; Huang, C.; Li, J. Z.; Grewe, B. F.; Zhang, Y.; Eismann, S.; Schnitzer, M. J., High-Speed Recording of Neural Spikes in Awake Mice and Flies with a Fluorescent Voltage Sensor. *Science* **2015,** *350*, 1361-1366.
5. Hochbaum, D. R.; Zhao, Y.; Farhi, S. L.; Klapoetke, N.; Werley, C. A.; Kapoor, V.; Zou, P.; Kralj, J. M.; Maclaurin, D.; Smedemark-Margulies, N.; Saulnier, J. L.; Boulting, G. L.; Straub, C.; Cho, Y. K.; Melkonian, M.; Wong, G. K.; Harrison, D. J.; Murthy, V. N.; Sabatini, B. L.; Boyden, E. S.; Campbell, R. E.; Cohen, A. E., All-Optical Electrophysiology in Mammalian Neurons Using Engineered Microbial Rhodopsins. *Nat. Methods* **2014,** *11*, 825-33.
6. Paget, V.; Sergent, J. A.; Grall, R.; Altmeyer-Morel, S.; Girard, H. A.; Petit, T.; Gesset, C.; Mermoux, M.; Bergonzo, P.; Arnault, J. C.; Chevillard, S., Carboxylated Nanodiamonds Are Neither Cytotoxic nor Genotoxic on Liver, Kidney, Intestine and Lung Human Cell Lines. *Nanotoxicology* **2014,** *8 Suppl 1*, 46-56.
7. Thalhammer, A.; Edgington, R. J.; Cingolani, L. A.; Schoepfer, R.; Jackman, R. B., The Use of Nanodiamond Monolayer Coatings to Promote the Formation of Functional Neuronal Networks. *Biomaterials* **2010,** *31*, 2097-2104.
8. Huang, Y.-A.; Kao, C.-W.; Liu, K.-K.; Huang, H.-S.; Chiang, M.-H.; Soo, C.-R.; Chang, H.-C.; Chiu, T.-W.; Chao, J.-I.; Hwang, E., The Effect of Fluorescent Nanodiamonds on Neuronal Survival and Morphogenesis. *Sci. Rep.* **2014,** *4*.
9. Hsu, T.-C.; Liu, K.-K.; Chang, H.-C.; Hwang, E.; Chao, J.-I., Labeling of Neuronal Differentiation and Neuron Cells with Biocompatible Fluorescent Nanodiamonds. *Sci. Rep.* **2014,** *4*.
10. Faklaris, O.; Joshi, V.; Irinopoulou, T.; Tauc, P.; Sennour, M.; Girard, H.; Gesset, C.; Arnault, J. C.; Thorel, A.; Boudou, J. P.; Curmi, P. A.; Treussart, F., Photoluminescent Diamond Nanoparticles for Cell Labeling: Study of the Uptake Mechanism in Mammalian Cells. *ACS Nano* **2009,** *3*, 3955.
11. Haziza, S.; Mohan, N.; Loe-Mie, Y.; Lepagnol-Bestel, A.-M.; Massou, S.; Adam, M.-P.; Le, X. L.; Viard, J.; Plancon, C.; Daudin, R.; Koebel, P.; Dorard, E.; Rose, C.; Hsieh, F.-J.; Wu, C.-C.; Potier, B.; Herault, Y.; Sala, C.; Corvin, A.; Allinquant, B.; Chang, H.-C.; Treussart, F.; Simonneau, M., Fluorescent Nanodiamond Tracking Reveals Intraneuronal Transport Abnormalities Induced by Brain-Disease-Related Genetic Risk Factors. *Nat. Nanotechnol.* **2017,** *12*, 322-328.
12. Balasubramanian, G.; Chan, I. Y.; Kolesov, R.; Al-Hmoud, M.; Tisler, J.; Shin, C.; Kim, C.; Wojcik, A.; Hemmer, P. R.; Krueger, A.; Hanke, T.; Leitenstorfer, A.; Bratschitsch, R.; Jelezko, F.; Wrachtrup,24